\documentclass[pre,aps,groupaddress,twocolumn,superscriptaddress]{revtex4-1}
\usepackage{epsfig,amsmath,graphicx,amssymb,overpic}
\usepackage{bm}
\usepackage{graphicx}
\usepackage{subfigure}

\def\be{\begin{equation}}
\def\ee{\end{equation}}
\def\bee{\begin{eqnarray}}
\def\ene{\end{eqnarray}}
\def\bes{\begin{subequations}}
\def\ees{\end{subequations}}

\newcommand{\bq}{{\bf q}}
\newcommand{\bc}{\mathbf{c}}

\newcommand{\bA}{\mathbf{A}}
\newcommand{\bB}{\mathbf{B}}

\newcommand{\tc}{\tilde{c}}

\newcommand{\ii}{\mathrm{i}}
\newcommand{\p}{{\cal P}}
\newcommand{\PT}{{\cal PT}}
\newcommand{\T}{{\cal T}}

\begin{document}

\title{Parity-time-symmetric vector rational rogue wave solutions \\ in any
  $n$-component nonlinear Schr\"odinger models}

\author{Guoqiang Zhang}
\affiliation{Department of Mathematical Sciences, Tsinghua University, Beijing 100084, China}

\author{Liming Ling}
\email{linglm@scut.edu.cn (corresponding author)}
\affiliation{School of Mathematics, South China University of Technology, Guangzhou 510640, China}

\author{Zhenya Yan}
\email{zyyan@mmrc.iss.ac.cn (corresponding author)}
\affiliation{Key Lab of Mathematics Mechanization, Academy of Mathematics and Systems Science, Chinese Academy of Sciences, Beijing 100190, China}
\affiliation{School of Mathematical Sciences, University of Chinese Academy of Sciences, Beijing 100049, China}

\author{Vladimir V. Konotop}
\affiliation{Departamento de F\'{i}sica,  Faculdade de Ci\^encias,
Universidade de Lisboa, Campo Grande, Edif\'icio C8, Lisboa 1749-016,
Portugal}
\affiliation{Centro de F\'{i}sica Te\'orica e Computacional, Universidade de Lisboa, Campo Grande, Edif\'icio C8, Lisboa 1749-016, Portugal.\\
{\rm (Date:\,\, \today)}
}


\begin{abstract}

 The extreme events are investigated for an $n$-component nonlinear Schr\"odinger ($n$-NLS) system in the focusing Kerr-like nonlinear media, which appears in many physical fields. We report and discuss the novel multi-parametric families of vector rational rogue wave (RW) solutions featuring the parity-time ($\p\T$) symmetry, which are characterized by non-identical boundary conditions for the components, and consistent with the degeneracy of $n$ branches of Benjamin-Feir instability. Explicit examples of $\p\T$-symmetric vector RWs are presented.
Some parameter constraints can make some components generate the RWs with high amplitudes due to many-body resonant interactions.
 Effect of a non-integrable deformation of the model on the excitation of vector RWs is also discussed. These results will be useful to design the RW experiments in  multi-component  physical systems.

\end{abstract}



\maketitle

{\it Introduction.}---Recently, rogue waves (RWs), as a special type of nonlinear waves, have been paid more and more attention to. RWs
(alias freak or killer waves) are extreme events that were originally known as deep ocean waves~\cite{RW1,RW2}  (see also Refs.~\cite{Rev1,Rev2,Rev3} and references therein).
Nowadays RWs are recognized to be a ubiquitous phenomenon that can be observed in nonlinear physical systems of diverse nature including optical fibers~\cite{prw,rw2010, Dudley2014,jop,Mihalache1,mat,orw}, arrays of waveguides~\cite{BluKonAkh}, cavities~\cite{oc}, resonators~\cite{Majid}, superfluids~\cite{superfluid}, Bose-Einstein condensates (BECs)~\cite{bec-rw,Bludov2010},
models of atmospheric physics~\cite{ap,Iafrati}, plasmas~\cite{Langmur,RW_plasma}, and even finance~\cite{yanfrw}. When RWs are characterized by a scalar field, a paradigmatic physical model describing them
is the NLS equation or its generalizations. If a system is characterized by two or more (say, $n$) components, like for example, multi-mode nonlinear waveguides~\cite{nls1b,nlsb2} or multi-component BECs~\cite{nls5}, the governing equations are often reduced to a  two- or $n$-component NLS system (alias $n$-NLS system)~\cite{nnls}.

It is well established that occurrence of RWs is intimately  related  to the phenomenon of the modulation instability (MI)~\cite{MI,Onorato,MI-nonlinear-MI,MI-nonlinear-MI-1,Bludov2010,Baronio,Dudley2014}.
A focusing scalar NLS equation admits one branch of the spectrum featuring MI (or briefly one branch of MI), which in its turn originates a single type of rational RWs, known as the Peregrine soliton~\cite{Peregrine1983}. Such a solution is localized both in time and in space~\cite{Majid2,prl2017}, and has been experimentally observed in nonlinear optical fibers~\cite{rw2010,Dudley2014}, plasma~\cite{RW_plasma}, and  a water  tank~\cite{RW_tank}.
Meantime, already 2- and 3-component NLS equations may possess several branches of MI. Since with each branch of MI one can relate a RW solution,  one can expect that in a system with several field components different branches of MI can give origin to distinct RWs. Such vector RWs in 2-- and 3--component systems were extensively studied~\cite{Onorato,yanfrw,Bludov2010,Baronio,Zhao2012, Kedziora2013, Baronio121318,yan-coupled,Mihalache2}. An $n$-NLS system generally admits $n$ distinct branches of MI. Since with each branch one can relate a RW, a natural issue arises, on whether $n$ distinct branches of MI could collapse, thus manifesting degeneracy of MI, and thus result in a generation of a RW with greatly enhanced amplitude. A general answer to this issue still remains open, although  a few examples
were reported~\cite{Zhao2012}.

In this paper, we would like to tackle the above-mentioned open and interesting issue in a general case, and for the first time find
the analytical $n$-vector RWs, addressing the relation between the degenerated MI and vector RWs in the $n$-NLS equation, in which one of key and difficult points approaching the problem requires to solve an $(n+1)$-th-order polynomial equation {\it explicitly}.
Furthermore, we find that the obtained $n$-vector RWs feature the physically interesting parity-time ($\p\T$) symmetry, resembling the scalar time-reversal RWs previously reported in \cite{time-revers,Chabchoub14}, and the vector RWs can generate the high amplitudes for some parameter choices.

\emph{The $n$-NLS models and existence of vector RWs.}---We consider the wave propagation of $n$ self-trapped beam in a slow media 
with Kerr-like nonlinearity, a dimensionless $n$-NLS system for a complex-valued $n$-component column-vector field $\mathbf{q}(x,t)=\left(q_1(x,t), \ldots, q_n(x,t)\right)^{\rm T}$~\cite{nnls,scott,yeh,nail98,nail99,Ablowitz2004}
  \begin{equation}
  \label{n-NLS}
  \mathrm{i}\frac{\partial{\bf q}}{\partial t}+\frac{1}{2}\,\frac{\partial^2{\bf q}}{\partial x^2}+|\mathbf{q}|^2\mathbf{q}=\mathbf{0},
  \end{equation}
where $|\mathbf{q}|^2=|q_1|^2+\cdots+|q_n|^2$ stands for the norm of the column-vector $\bq$ and the superscript {\small T} stands for transpose.  At $n=1$ Eq.~(\ref{n-NLS}) is reduced to the scalar NLS equation, while at $n=2$ it is the well-known Manakov system~\cite{Manakov}. Eq.~(\ref{n-NLS}) also has direct physical applications, for instance, in the theory of  alpha-helixes~\cite{scott}.
Here we are interested in vector RW solutions of Eq.~(\ref{n-NLS}) propagating against a constant-amplitude carrier-wave (CW) background. The latter is considered of a general form $\mathbf{q}_0(x, t)=(q_{01},\ldots, q_{0n})^{\rm T}$, where $q_{0j}=a_j \mathrm{e}^{\mathrm{i}\varphi_j}$, $\varphi_j=k_j x-\omega_jt$ are the real phases, and positive amplitudes $a_j$, (real) wave-numbers $k_j$, and frequencies $\omega_j$ satisfy the dispersion relations: $\omega_j=k_j^2/2-|\mathbf{a}|^2$ with $\mathbf{a}=(a_1, \ldots, a_n)^\mathrm{T}$. The components of the RWs, $q_j$, are all localized (against the CW background) in space and in time and at $|x|,\, |t|\to\infty$ they have the asymptotics: $q_j\sim q_{0j}{\rm e}^{-2\mathrm{i}\theta_j}$ with constant phases $\theta_{j}$.

To present the $n$-vector RWs~\cite{Suppl} $q_j(x,t)=q_{0j}-4\mathrm{i}{\rm Im}(\lambda_0)\!\left(\Psi\Psi^{\dag}/(\Psi^{\dag}\Psi)\right)_{j+1,1}$ with `$\dag$' being the Hermitian conjugation via the Darboux transform (see e.g.,~\cite{dt,Akhmediev2009a,Guo2012,he2013}), one needs to find the auxiliary vector function of the Lax pair~\cite{Ablowitz2004}: $\Psi=\mathbf{G}\Phi\mathbf{c}$, $\mathbf{G}=\mathrm{diag}(1, \mathrm{e}^{\mathrm{i}\varphi_1}, \ldots, \mathrm{e}^{\mathrm{i}\varphi_n})$, the non-zero constant column-vector ${\bf c}=(c_0,c_1,\ldots, c_n)^{\rm T}$ introduces free parameters $c_j$ (as shown below they characterize RWs), and $\Phi$ is an $(n+1)$-order square matrix solution of the gauge-transformed Lax pair with constant coefficients:
$\Phi_x=\mathrm{i}\mathbf{H}\Phi$ and $\Phi_t=\mathrm{i}\left[\frac12\mathbf{H}^2+\lambda_0\mathbf{H}-\left(|\mathbf{a}|^2+\frac12\lambda_0^2\right)\mathbb{I}_{n+1}\right]\Phi$, where
 $\mathbf{H}=\lambda_0\sigma_3+\begin{pmatrix}
		0 & \mathbf{a}^{\rm T}  \\
		\mathbf{a} &-\mathbf{K}
	\end{pmatrix}
	$ with $\mathbf{K}=\mathrm{diag}\left(k_1,\cdots, k_n\right)$ being a constant matrix, and $\lambda_0\in\mathbb{C}$ the spectral parameter.
Thus, the construction of a desired solution $\bq(x,t)$ is reduced to solving the eigenvalue problem. Representing (for convenience) an eigenvalue of ${\bf H}$ as ${\rm i}\chi-\lambda_0$, i.e. regarding $\chi$ as an eigenvalue, one obtains the characteristic equation for ${\bf H}$ in the form~\cite{Suppl}
\bee\label{ce}
P(\chi):=2\lambda_0 -{\rm i}\chi+\sum_{j=1}^{n}\frac{a_j^2}{{\rm i}\chi+k_j}=0.
\ene

We are interested in vector {\em rational} RWs of Eq.~(\ref{n-NLS}), which are represented by a quotient of two polynomials of $x$ and $t$ [like in Eqs.~(\ref{components1}) and (\ref{components2}) shown below]~\cite{Peregrine1983}. Therefore we seek matrix solutions of the above-mentioned equations
that feature {\em algebraic} dependence on the spatial (and temporal) variables. This implies that roots of Eq.~(\ref{ce}) determining such solutions must have multiplicity bigger than one. Finding such roots is a nontrivial task because effectively Eq.~(\ref{ce}) is an algebraic equation of order $(n+1)$, whose roots of high multiplicity may even do not exist in a general case. Furthermore, here we consider the roots of largest possible, i.e.,  $n+1$ multiplicity. Thus, in addition to Eq.~(\ref{ce}) we require
\begin{equation}
\label{eq:mi}
P^{(s)}(\chi):=\delta_{s1}+\sum_{j=1}^{n}\frac{a_j^2}{({\rm i}\chi+k_j)^{s+1}}=0,
\end{equation}
where $s=1,\ldots,n$ and $\delta_{ij}$ is the Kronecker delta. Since straightforward obtaining $\chi$ from Eqs.~(\ref{ce}) and (\ref{eq:mi}) seems to be impossible, we reformulate the problem and focus on finding the non-zero amplitudes $a_j$ and wave-numbers $k_j$ of the CW background, for which a given value of the spectral parameter $\chi=\chi_0$ is the root of the multiplicity $n+1$, i.e.,  $P(\chi_0)=P^{(1)}(\chi_0)=\cdots=P^{(n)}(\chi_0)=0$. We notice that  $P^{(1)}(\chi)=0$ is the dispersion relation for the MI~\cite{ling2019} and for $n=2,3$ it was shown~\cite{Zhao2012} to coincide with the condition for the baseband MI~\cite{Baronio,Baronio121318} of
Eq.~(\ref{n-NLS}).

Due to the Galilean invariance of Eq.~(\ref{n-NLS}) the $\chi_0$ can be made real
by the 
shift of the wave-numbers $k_j\to k_j -{\rm Im}(\chi_0)$. Furthermore, by rescaling  $a_j$ and $k_j$  one can make $\chi_0=1$. 
Seeking the solutions of Eqs.~(\ref{ce}) and (\ref{eq:mi}) that are located on the left branch of the hyperbola: $a_j^2-k_j^2=1$, we obtain
\bee \label{parameters}
 a_j=\csc\theta_j,\, k_j=\cot\theta_j, \, \theta_j=\frac{ \pi j}{n+1},\, \lambda_0=\frac{{\rm i}}{2}(n\!+\!1).\,\,
 \ene
Thus, we have found the phases $\theta_j$ of the components, which determine the amplitudes $a_j$, the wave-numbers $k_j$, and the frequencies $\omega_j$ of the CW background sustaining a vector rational RW. The analytic expression for the respective RW is obtained by computing
 $\Phi$ which corresponds to the $(n+1)$-multiple eigenvalue $\ii\chi_0-\lambda_0={\rm i}(1-n)/2$ of ${\bf H}$, and its substitution into the Darboux transform.
The fundamental solution determined by $\Phi(0,0)=\mathbb{I}_{n+1}$, and hence represented as $\Psi={\bf G}\Phi{\bf c}$, generates a family of vector raiotnal RWs~\cite{Suppl}
\begin{equation}
\label{1-RW}
q_j(x,t)=q_{0j}\left[ 1-
\frac{2\mathrm{i}\left(n+1\right)}{a_j}\frac{\mathbf{A}_{j+1}{\bf c}(\mathbf{A}_1{\bf c})^*}
{(\mathbf{A}{\bf c})^\dag \mathbf{A}{\bf c}}\right],
\end{equation}
where the asterisk denotes the complex conjugation, $\mathbf{A}_{j}$ is the $j$-th row of the matrix function
\begin{equation}
\label{A}
\mathbf{A}\!=\!\sum_{s=0}^n\sum_{m=0}^{\lfloor n/2\rfloor}\!
\frac{(\mathrm{i} x\!-\!t)^s(\mathrm{i} t)^m }{2^m s!\,m !} \mathbf{B}^{s+2m}, \,\,
\mathbf{B}\!=\!\begin{pmatrix} {\rm i}n &\mathbf{a}^{\rm T}\\ \mathbf{a} & -{\rm i}\mathbb{I}_n\!-\!\mathbf{K}\end{pmatrix}
\end{equation}
and $\lfloor\cdot\rfloor$ stands for the integer part.


\emph{$\PT$-symmetric vector rogue waves.}---Generally speaking, the obtained vector RW solution (\ref{1-RW}) does not inherit the symmetry of the CW background (\ref{parameters}) that is symmetric with respect to simultaneous space inversion  $x\to-x$ and swapping  the components performed by the matrix $\p=(\delta_{j, n+1-j'})_{n\times n}$. Meantime, it is possible to impose a constraint on the parameters ensuring the $\PT$-symmetry of a vector RW, with $\T$ being the conventional bosonic time-reversion: $t\to -t,\, {\rm i}\to -{\rm i}$. Indeed, the formula (\ref{1-RW}) yields a $\PT$-symmetric vector RW: ${\bf q}(x,t)={\cal PT}{\bf q}(x,t)={\cal P}{\bf q}^*(x,-t)$ if $\mathbf{c}$ is required to satisfy the symmetry: ${\bf c}=\widehat{\cal P}{\bf c}^*$ with $\widehat{\cal P}=\begin{pmatrix}1&\mathbf{0}_{1\times n} \\ \mathbf{0}_{n\times 1} & - {\cal P} \end{pmatrix}$.

By construction $\bB$ is a defective matrix with the $n$-th order exceptional point for the parameter choices (\ref{parameters}). Therefore, it is convenient to define a set of generalized eigenvectors
 $\mathbf{s}_{\ell}=(\delta_{\ell 0}, {a_1}/{\left({\rm i}+k_1\right)^{\ell+1}}, \ldots, {a_n}/{\left(\mathrm{i}+k_n\right)^{\ell+1}})^\mathrm{T}$, $\ell=0,\ldots, n$,
satisfying $\mathbf{B}^\ell\mathbf{s}_\ell\ne \mathbf{0}$ and $\mathbf{B}^{\ell+1}\mathbf{s}_\ell=\mathbf{0}$.
 The vector $\bc$
can be spanned over the basis $\mathbf{S}=\left(\mathbf{s}_0,
\ldots, \mathbf{s}_n\right)$:  $\mathbf{c}=\mathbf{S}\boldsymbol{\alpha}$, where the form of the expansion coefficient, compactly represented by the column-vector $\boldsymbol{\alpha}$, must be chosen to ensure the validity of ${\bf c}=\widehat{\cal P}{\bf c}^*$. This is achieved by $\boldsymbol{\alpha}=\left(\varepsilon_0\alpha_0,
\ldots, \varepsilon_n\alpha_n\right)^\mathrm{T}$ with all $\alpha_{\ell}\in\mathbb{R}$, $\varepsilon_{\ell}=1$ for even ${\ell}$ and $\varepsilon_{\ell}={\rm i}$ for odd ${\ell}$. Furthermore, since one of nonzero parameters $\alpha_{\ell}$ can be arbitrarily fixed, we set $\alpha_{\ell}=1$ when $\alpha_{\ell+1}=\cdots=\alpha_n=0$. With this  choice  $\mathbf{Ac}$ is a vector polynomial of  degree $\ell$.

Resuming, a family  of $\PT$-symmetric vector RWs (\ref{1-RW}) obtained against a given CW background with parameters (\ref{parameters}), can be characterized by a pair of indexes $(n,\ell)$.
 The internal structure of such a RW is also characterized by the real parameters $\alpha_0,\ldots,\alpha_{\ell-1}$. Due to the translational invariance, one of the parameters $\alpha_j$ is a trivial spatial
shift determining the location of the RW event and hence, it can be scaled out by the
choice of the system of coordinates. Other $\ell-1$ parameters determine the structure of the RW, as it is discussed below.

\begin{figure}[!t]
	\centering
	\includegraphics[width=\columnwidth]{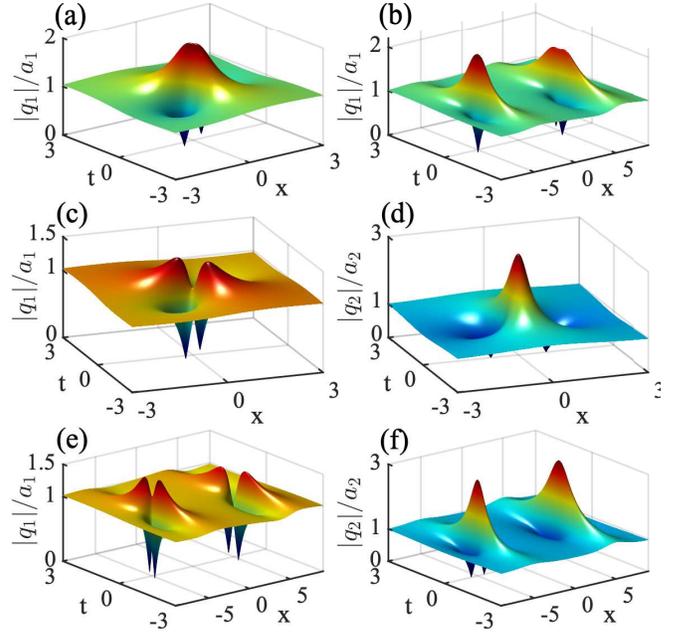}
\caption{Profiles of the RW components~(\ref{components1}) (panels a, c, d) and (\ref{components2}) (panels b, e, f) for the cases: (a) $(n, \ell)=(2,1)$;
(b) $(n, \ell)=(2,2)$ with $\alpha_0=12.5$; (c) and (d) $(n, \ell)=(3,1)$;
		(e) and (f) $(n, \ell)=(3,2)$ with $\alpha_0=12.5$.
	}
	\label{fig1-rw}
\end{figure}

\emph{Examples of vector rational rogue waves.}---The fundamental vector rational RWs correspond to $\ell=1$ (respectively $\alpha_\ell=1$) in (\ref{1-RW}). Choosing  $\alpha_0=-1/2$  we obtain:
\begin{gather}
\label{components1}
q_j(x,t)=\left(a_j+\frac{\mathrm{i}}{a_j}\frac{2( k_jx-t)+\mathrm{i}}
 {x^2+ t^2+ 1/4}\right){\rm e}^{\mathrm{i}(k_jx-\omega_jt)-2\mathrm{i}\theta_j}.
\end{gather}
The $(n,1)$-RW solution has no free parameter and is fully determined by the characteristics of the CW background.
A number of humps and dips, i.e. of the most interesting events, in each component depend on the number of components~\cite{Suppl}. At $n=1$ we recover the Peregrine soliton having one hump and two dips~\cite{Peregrine1983}. For $n=2$, when $\theta_1=\pi/3$ and $\theta_2=2\pi/3$ [see Eq.~(\ref{parameters})], the RW components are related by the $\PT$-transform, and now each component has one hump and two dips as this is illustrated for $n=2$ in Fig.~\ref{fig1-rw}(a).
For $n\geq3$ the components with $j\in \left(0, (n+1)/6\right]\cup [5(n+1)/6, n]$ have two humps and one dip (for this case to take place one must have at least $n=5$ components); the components with $j\in \left((n+1)/6, (n+1)/3\right)\cup \left(2(n+1)/3, 5(n+1)/6\right)$ have two humps and two dips [see Fig.~\ref{fig1-rw} (c)], the components with $j\in \left[(n+1)/3, 2(n+1)/3\right]$ have one hump and two dips [see Fig.~\ref{fig1-rw} (d)].

Next we consider examples of $(n,2)$-RWs, whose spatial location is parameterized by $\alpha_1$. We scale it out by the choice $\alpha_1= 1/2$. However now there exists one free parameter $\alpha_0$, which determines the shape of the  RWs components. From (\ref{1-RW}) we obtain
\begin{eqnarray}
\label{components2}
q_j(x,t)=\left(a_j+\frac{\mathrm{i}}{a_j}\frac{2r_{0j}-(4\alpha_0+1/2) r_{1j}+k_j+\mathrm{i}}{r +({x}^2-t^2-2\alpha_0-1/4)^2}\right)
\nonumber\\
\times {\rm e}^{\mathrm{i}(k_jx-\omega_jt)-2\mathrm{i}\theta_j},\quad
\end{eqnarray}
where $r=[(2{x}-1)^2+1]t^2+({x}+1/2)^2+1/4$, $r_{0j}=2(k_j{x}-t)({x}^2+t^2)+({\rm i}-2k_j)t^2+k_jx
+[(3\mathrm{i}k_j-1){x}^2+4k_j{x}t-2t^2-2{x}-2(k_j-\mathrm{i})t]/(k_j+\mathrm{i})$
and $r_{1j}=2(k_j{x}+t)+(\ii k_j+1)/(k_j+\mathrm{i})$. Fig.~\ref{fig1-rw}(b) displays
the fundamental (2,2)-RW with one hump and two dips, while Figs.~\ref{fig1-rw}(e) and (f) exhibit the components of a (3,2)-RW. We observe that the second component is characterized by ``dominating" humps while in the first components the dips are more pronounced.

\emph{Polymeric $\PT$-symmetric vector rational RWs.}--- The diversity of multi-component $\PT$-symmetric vector RWs (\ref{1-RW}) at large $n$ is abundant. Indeed, for a given $n$ one has $\ell=1,\ldots,n$ different solutions characterized by $\ell-1$ nontrivial free parameters, and hence $n(n-1)/2$ distinct types of RWs.  One can expect that the strongest collective effect of the components is achieved when they all acquire their maxima at the same instant of time and spatial location. Without loss of generality we can look for such a solution with the largest hump occurring at $t=0$ and $x=0$, when $\bA(0,0)=\mathbb{I}_{n+1}$ [see (\ref{A})]. Since one of the components of $\bc$ can be arbitrarily chosen, we can set $c_0=1$ and $c_j=\ii \tc_j$ for $1\leq j\leq n$, where real $\tc_j$ are to be determined. Now we deduce from the general formula (\ref{1-RW}) that the maximum of $|q_j(0,0)|$ is achieved if the second term in the square brackets is real negative, and hence $\tc_j$ are all real positive. In that case $q_j(0,0) =a_j+\eta_j$ with $\eta_j=2(n+1)\tc_{j}/(1+|\tilde{\bf c}|^2),\,\, \tilde{\bf c}=(\tc_1,\ldots,\tc_n)^{\rm T}$
is the deviation of the amplitude of the $j$-th component from the background.

	
\begin{figure}[!t]
		\centering
		\includegraphics[width=\columnwidth]{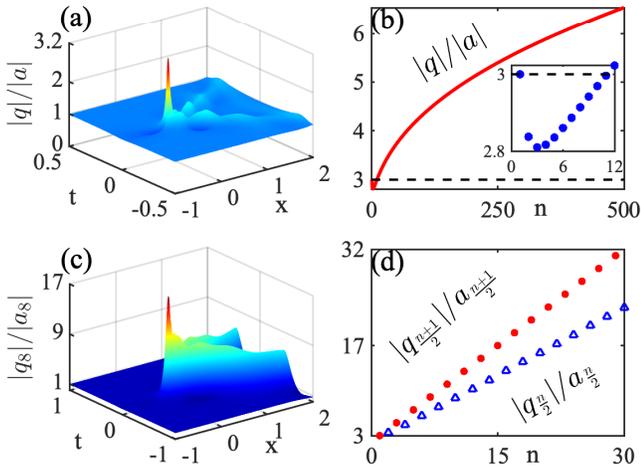}
	\vspace{-0.2in}	\caption{ (a, b) Profiles $|{\bf q}(x,t)|/|{\bf a}|$ of RW (\ref{1-RW}) for ${\bf c}=(1, \mathrm{i}/\sqrt{n}, \ldots, \mathrm{i}/\sqrt{n})^{T}$ and (a) $n=15$ or (b) $n\in [1, 500]$. (c, d) The profiles of RW (\ref{1-RW}) for ${\bf c}=(1, 0,\cdots,0,{\rm i},0,\cdots,0)^T$ with odd $n$ and ${\bf c}=(1,0,\cdots,0,{\rm i}/\sqrt{2},{\rm i}/\sqrt{2},0,\cdots,0)^T$ with even $n$:
(c) $|q_8(x,t)|/a_8$ with the ultrahigh peak amplitude ${\rm max}(|q_8|/a_8)=17$ for $n=15$; (d) The curves of ${\rm max}(|q_{j}(x,t)|/a_{j})=|q_{j}(0,0)|/a_{j}$ for $j=(n+1)/2$ with odd $n$ and $j=n/2$ with even $n$ as $n\in [1,30]$. }
 \label{fig:huge}
	\end{figure}

{\it Case A.}-For the parameter ${\bf c}=(\sqrt{n}, \mathrm{i}, \mathrm{i},\cdots, \mathrm{i})^{T}$, the average amplitude $\left|\mathbf{q}(x, t)\right|\left/\right.\left|\mathbf{a}\right|$ of the polymeric RWs attain maximum at the origin, where $\left|\mathbf{q}(x, t)\right|=\sum_{s=1}^n\left|q_s(x, t)\right|$ and $\left|\mathbf{a}\right|=\sum_{s=1}^n a_s$. In this case,  the ratio between peak value $|\mathbf{q}(0,0)|$ and CW background $|\mathbf{a}|$ can be deduced as
\begin{equation}
Ra=\frac{|\mathbf{q}(0,0)|}{|\mathbf{a}|}=1+\frac{\sqrt{n}(n+1)}{\sum_{j=1}^n\csc\theta_j},
\end{equation}
which has the following estimates
\begin{equation}\label{guji}
1+\frac{\sqrt{n}}{\ln\left(n+1\right)+1}\le Ra \le 1+\frac{\sqrt{n}\pi}{2\ln(n/2)},
\end{equation}
which implies $Ra\leq 3$ for $2\leq n\leq 11$ and $Ra\gg3$ for $n\gg 12$, and $Ra=\mathcal{O}\left(\sqrt{n}/\ln n\right)$ as $ n\to \infty$.
Fig.~\ref{fig:huge}(a) shows the distribution structure of $\left|\mathbf{q}(x, t)\right|\left/\right.\left|\mathbf{a}\right|$ for $15$-NLS equations. Fig.~\ref{fig:huge}(b) indicates the ratio $|\mathbf{q}(0,0)|/|\mathbf{a}|$ for $n$. Besides, the ratio between peak value and background for the $j$-th component is $|q_j(0,0)|/a_j=1+(n+1)/\sqrt{n}\sin\theta_j$.

{\it Case B.}-In this case, we consider the polymeric RW, whose middle component attains maximum at the origin. Due to the restrictive of the time-reversal property, two categories are studied respectively. As $n$ is odd, as long as $\mathbf{c}=\left(1, 0, \cdots, 0, \mathrm{i}, 0,\cdots, 0\right)$, the amplitude of $(n+1)/2$-th component attains maximum $q_m(0,0)/a_m=(n+2),\, m=(n+1)/2$ at the origin,
and others are the height $a_j's$ of  the backgrounds at the origin. Fig. \ref{fig:huge}(c) shows the eighth  component of polymeric RWs for $15$-NLS equations. The red dots in Fig. \ref{fig:huge}(d) shows the values of ${q_m(0,0)}/{a_m}$ with $m=(n+1)/2$ and odd $n\in [1, 30]$.
As $n$ is even, as long as $\mathbf{c}=\left(1, 0, \cdots, 0, \mathrm{i}/\sqrt{2}, \mathrm{i}/\sqrt{2}, 0,\cdots, 0\right)$, the amplitudes of $n/2$- and $(n+1)/2$-th components attain maximum at the origin with $m=n/2$ and
$q_{m}(0,0)/a_m=q_{m+1}(0,0)/a_{m+1}=1+(n+1)/\sqrt{2}\sin(\theta_m)$, and others are the height $a_j's$ of the corresponding backgrounds. The blue triangles in Figs. \ref{fig:huge}(d) show the values of ${q_m(0,0)}/{a_m}$ with $m=n/2$ and even $n\in [1, 30]$.

	\begin{figure}[!t]
	\centering
	\includegraphics[width=\columnwidth]{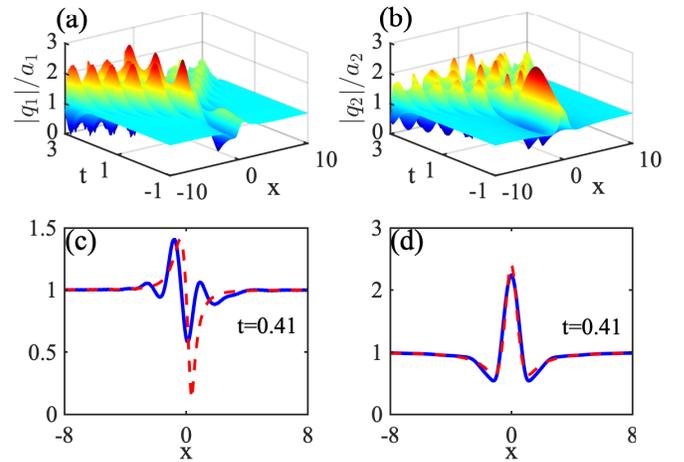}
	\vspace{-0.2in}
	\caption{Numerical simulations of non-integrable 3-NLS equation (\ref{3-nls}) with $\delta=0.05$. Upper panes show evolution of $|q_j|/a_j\, (j=1,2)$ two non-equal components (recall that due to the $\PT$-symmetry $|q_3|=|q_1|$).  Bottom panels show comparison of numerical solutions at $t=0.41$ (blue solid curves)  and exact RWs (red dashed curves) corresponding to $\delta=0$, i.e., given by Eq.~(\ref{components1}) [cf. with Figs.~\ref{fig1-rw}(c) and (d)].}
\label{fig:numer}
\end{figure}

{\it Non-integrable $n$-NLS model.}---In a general setting a multi-parametric non-integrable system
 cannot be solved analytically. Even numerical study would require particularization depending on the physical nature of the nonlinearity. Therefore here we address a somewhat simpler question whether $\PT$-symmetric RWs of enhanced intensity can be excited in an $n$-NLS equation,  where the integrability is broken by unequal nonlinear interactions. We explore this issue by simulating evolution of vector RWs in the dimensionless non-integrable 3-NLS equation~\cite{nail98,nail99}:
 \begin{equation}
 \label{3-nls}
 {\rm i}\frac{\partial q_{j_1}}{\partial t}+\frac12\frac{\partial^2q_{j_1}}{\partial x^2}+\left[|q_{j_1}|^2+(1-\delta)(|q_{j_2}|^2+|q_{j_3}|^2)\right]q_{j_1}=0,
 \end{equation}
 where $j_1$, $j_2$, and $j_3$ are all different and acquire values $1$, $2$ and $3$, and $\delta$ is a small real parameter quantifying imbalance between intra- and inter-mode interactions. However, in spite of the $\delta$-deformation, model (\ref{3-nls})  remains $\p$-symmetric, i.e., is not affected by swapping the components. For numerical simulations via the Fourier spectral method we consider small deviation from the integrable limit letting $\delta=0.05$ and propagate the ``initial" condition, in the form of the exact $\PT$-symmetric vector RW (\ref{components1}) taken at  $t=-1$ [Figs.~\ref{fig:numer} (a) and (b)]. Although the post-extreme-event evolution of the non-integrable model is appreciably different from its integrable counterpart [cf. with Figs.~\ref{fig1-rw} (c) and (d)], the RW itself is well distinguishable in the evolution.  In Figs.~\ref{fig:numer} (c) and (d) we compare the numerical solutions of (\ref{3-nls}) with the exact RW solutions (\ref{components1}) at the instant of occurrence of the maximum, i.e., at $t=0.41$ in our case. We observe that the structures and especially the shapes of the pick amplitudes are very close in both models.

{\it Conclusions.}---To conclude, we described a family of exact $\PT$-symmetric vector rational RW solutions of an integrable system of $n$-component coupled NLS equations.  Such extreme events occur against CW backgrounds which are characterized by different asymptotics of different components. The higher amplitudes reached during the evolution increase with the number of the components. The reported RWs are observable in non-integrable systems, too. The obtained vector RW solutions may pave the way for obtaining high-amplitude RWs in diverse multi-component physical systems including multi-mode and poly-chromatic optical devices, as well as multi-atomic and spinor Bose-Einstein condensates. Moreover, we also
give the higher-order vector RWs and asymptotic estimates of the $n$-NLS system in another literature~\cite{long}. The idea can also be extended to other $n$-component nonlinear integrable systems (e.g., $n$-component Hirota equations, mKdV equations, complex mKdV equations, higher-order NLS equations, KP equations, and etc.)~\cite{long}.  Being restricted to all attractive interactions, our study left open such questions as obtaining high-intensity RWs in presence of attractive interactions, as well as establishing direct links between $\PT$-symmetric vector RWs and the modulation instability of the governing system.

{\it Acknowledgments.}-G.Z. acknowledges support from China Postdoctoral Science Foundation (Grant No.2019M660600). L.L. acknowledges support from the National Natural Science Foundation of China (Grant No.11771151), the Guangzhou Science and Technology Program of China (Grant No.201904010362), and the Fundamental Research Funds for the Central Universities of China (Grant No.2019MS110).  Z.Y. acknowledges support from the the National Natural Science Foundation of China (Grant Nos.11925108 and 11731014).  V.V.K. acknowledges financial support from the Portuguese Foundation for Science and Technology (FCT) under Contract no. UIDB/00618/2020.

\end{document}